\documentclass[article]{elsarticle}

\usepackage{hyperref,amsmath,amssymb,amsfonts,amsthm,algorithmic,algorithm}
\theoremstyle{plain}
\newtheorem{tw}[]{Theorem}
\theoremstyle{remark}

\theoremstyle{plain}

\usepackage{todonotes,subfig}

\usepackage{color}

\journal{arXiv}

\begin{document}

\begin{frontmatter}

\title{Bayesian inference for age-structured population model of infectious disease with application to varicella in Poland}

\author[impan,mimuw]{Piotr Gwiazda}\fnref{pg}
\author[mimuw]{B{\l}a\.zej Miasojedow}
\author[pzh]{Magdalena Rosi\'nska}\fnref{mr}
%
%
%
\address[impan]{Institute of Mathematics, Polish Academy of Sciences}
\address[mimuw]{Faculty of Mathematics, Informatics and Mechanics, University of Warsaw}
\address[pzh]{National Institute of Public Health - National Institute of Hygiene, Warsaw}
\fntext[pg]{Piotr Gwiazda received funding from the National Science Centre (Poland) grant 2015/18/M/ST1/00075.}
\fntext[mr]{Magdalena Rosi\'nska received funding from the National Science Centre (Poland) grant
DEC-2012/05/E/ST1/02218
}

\begin{abstract}
The dynamics of the infectious disease transmission are often best understood by taking 
into account the structure of population with respect to specific features, for example age or immunity level. 
The practical utility of such models depends on the appropriate calibration with the observed data. 
Here, we discuss the Bayesian approach to data assimilation in the case of a two-state age-structured model. 
Such models are frequently used to explore the disease dynamics (i.e. force of infection) 
based on prevalence data collected at several time points. 
We demonstrate that, in the case when the explicit solution to the model equation is known, 
accounting for the data collection process in the Bayesian framework allows us to obtain an unbiased posterior distribution for 
the parameters determining the force of infection. 
We further show analytically and through numerical tests 
that the posterior distribution of these parameters is stable with respect to a cohort approximation (Escalator Boxcar Train) of the solution. 
Finally, we apply the technique to calibrate the model based on observed sero-prevalence of varicella in Poland. 
\end{abstract}


\end{frontmatter}


\section{Introduction}
Application of mathematical modelling of natural phenomena has proved to be very useful in many areas 
including population dynamics and transmission of infectious diseases. 
The practical value of such models depends heavily on the assumptions  made while developing the model are realistic, 
whether it also depends on the assimilation of real data into the model to inform of model parameters.

The populations, which are heterogeneous with respect to some individual property, 
are often described with nonlinear first order hyperbolic equations (structured population models). 
In the models describing the epidemic processes in human population, examples of such parameters may include age, 
time from infection or the level of immunity induced by  past infection or vaccination. 
For example, evolving demographic structure has an impact on infectious disease transmission. 
It has been observed that the long term evolution of the dynamics of infectious disease is highly dependent on demographic 
transitions; a change of age structure changing from a young population model an to aging model is typical for developed counties.

The classical model of infectious diseases was introduced by Kermack and McKendric \cite{Kermack1991} 
and  variations og it also were studied  by e.g. Reddingius 
\cite{Reddingius1971}, Metz \cite{Metz1978}, Iannelli \cite{Iannelli1995}, 
Diekmann \cite{Diekmann1977, Diekmann2013}, Thieme \cite{Thieme1977}. 
These models consider, for example, variable infectivity and variable susceptibility to infection. 
In particular, the infectivity often depends on the time from infection and susceptibility to infection - on 
immunity acquired from past infection waning in time. 

A similar model, but with an age structure instead of time-since-infection, was considered by \cite{Iannelli1995, Diekmann2013}. \\ 
If only two states are considered, i.e. the susceptible and those who have ever been infected, the model simplifies to:
\begin{equation}\label{eqn0}
\partial_t q(t,a) + \partial_a q(t,a) = -\lambda(t,a) q(t,a) \ {\rm for} \ (t,a) \in \mathbb R \times \mathbb R^+\,.
\end{equation}
In this model $q(t,a)$ represents the proportion of susceptible individuals of age $a$ at a time point $t$. If we assume that all individuals are susceptible at birth this equation may be supplemented with boundary condition:
\begin{equation} \label{init0} 
q(0,t)=1\,\,\,\,\, {\rm for \,\, all} \,\,\, t \in \mathbb R\,.
\end{equation}
Note that in this problem no initial condition is needed.

This simple model has received particular attention due to its usefulness in epidemiological applications. 
It captures the situation when the disease occurs with age and time dependent 
frequency $\lambda$, but in an individual we are only able to distinguish whether or not  the disease has already occured. 
The model was applied to infectious diseases e.g. toxoplasmosis \cite{Aedes1993, Marschner1997,Marschner1996}, HIV \cite{Hallett2008, Mahiane2012}, hepatitis A \cite{ Keiding1991, Shkedy06}, rubella, mumps, varicella \cite{Shkedy06}, tuberculosis \cite{Nagelkerke1999} and non-infectious diseases, e.g. diabetes \cite{ Keiding1991, Bell2013}, myasthenia gravis \cite{ Keiding1991} or dementia \cite{Brinks2015}. 
In the case of infectious diseases which confer long lasting immune response, 
a marker of past or ongoing infection can be found, i.e. the measurable serum level of antibodies. 
Seroprevalence studies are quite often performed and these data can be assimilated into this model for further applications, including simulation studies and predictions. 
The parameter $\lambda(t,a)$ itself 
may be of interest describing a quantity which is often difficult to measure directly, but is important from epidemiological point of view; the force of infection. 
So far, the estimation methods for $\lambda(t,a)$ rely on maximizing an appropriately constructed likelihood function as outlined in \cite{NielHens2012}. 
Construction of the data model required both evaluation of the solutions of equation \eqref{eqn0} as well as the ability to account for the data aggregation process. 
The approach proposed in prior studies involves consideration of equation \eqref{eqn0} on characteristics, i.e. birth cohorts $(b+a, a)$, where b represents the time of birth and a is the age. With such parametrisation equation (\ref{eqn0}) can be rewritten as:

\begin{equation}\label{eqn1}
\frac{dq_b(a)}{da}=-l_b(a)q_b(a)\,,
\end{equation}
where $q_b(a)=q(b+a,a)$ and $l_b(a)=l(b+a,a)$.\\
Equation \eqref{eqn1} can be solved for $l_b(a)$:
\begin{equation} \label{eqn2}
l_b(a)=\frac{-q_b'(a)}{q_b(a)}=\frac{\pi_b'(a)}{1-\pi_b(a)}\,,
\end{equation}
where $\pi_b(a)=1-q_b(a)$ denotes the prevalence of the antibodies.\\

Considerable research has been carried out to find a flexible method of modelling $l(t,a)$, witch work on both  parametric and non-parametric approaches. 
These are usually based on factorization: $l(t,a)=l^1(t)\;l^2(a)$. Additionally, they are based on a form of $l(t,a)$, which allows the 
construction of a general linear model for $\pi(t,a)$, which in turn would be estimable from the available data \cite{NielHens2012}. 
The problem of data aggregation is tackled either by assuming a piecewise constant force of infection on age-time boxes relevant for the data 
resolution \cite{Hallett2008,Shkedy06}, or using a mid-point value of the solution on the characteristic for the aggregation interval, as in \cite{Brunet1999}.

In this paper we propose a Bayesian approach to estimate the equation parameter $\lambda$ based on available data. The Bayesian approach offers a flexible way to recover the 
full posterior distribution over the parameter, avoiding difficulties of estimating  confidence intervals, through  error propagation techniques. 
Acknowledging that previously the cohort formulation was commonly used in applications, 
we show how the approximation of the continuous case by the cohorts is reflected in the distance between the posterior distributions of the parameters. 
This distance depends in general on how densely the population is divided into the birth cohorts. 
When only few cohorts are considered, which has been  the case in applications so far, 
this can lead to considerable bias in the posterior distribution in the continuous case. 

We further note that  equation \eqref{eqn0} is a special case of structured population models. These models are often used in theoretical biology for a wide variety of models including evolution of populations, infectious diseases or cellular growth, see e.g. \cite{Iannelli1995,Roos2013,MR2914262}.
For the simple model defined by equation \eqref{eqn0} with boundary condition \eqref{init0} it is possible to find an explicit formula for the solution and 
employ it directly in the Bayesian inverse problem. Thus, the birth cohort 
approach can be viewed just as an alternative way of modelling the process. 
However, our aim is not only to tackle this particular problem but also to propose a general method, which could be applied for general structured population model.
For example, if we want to extend the model \eqref{eqn0} to incorporate vertical transmission, we should include more complicated boundary conditions, e.g.:
\begin{equation}\label{init1}
q(t,0)=\int\limits_0^\infty \beta(a) [1-q(t,a)]{ d}a\,.
\end{equation}
In this case we cannot solve this system explicitly and therefore we have to rely on an approximation scheme. 
For this aim, let us recall a recently developed framework for the analysis of structured population dynamics in the spaces of nonnegative 
Radon measures with a suitable metric which provides a rigorous tool to study numerical approximations of the system. 
One example of a such numerical algorithm widely applied in theoretical biology is Escalator Boxcar Train (EBT) \cite{Roos1988}. 
The approach is based on the idea of tracing growth and the transport of measures which approximate
the solution of the original partial differential equation. These measures are defined as sums of Dirac measures, each one of which represents the average state and number of individuals within a specific group. In terms of population studies the concept corresponds to following birth cohorts over time. We remark that when applying this technique in a Bayesian inverse problem it is possible to find an approximate posterior measure for the equation parameter. 

The distance of this measure from the posterior measure for the original problem will be related to the error of the EBT or similar particle approximation, depending on the model equation.  
In the recent papers (\cite{Carrillo2012, MR2746205,MR2644146}), theoretical results on the stability of the solution (stability of the semigroup) to the general form of the structured population equation 

\begin{equation*}
      \partial_t \mu
      +
      \partial_x \left( b(t, \mu) \, \mu \right)
      +
      c(t, \mu) \, \mu
      = 
      \int_{\mathbb {R}^+} \left(\eta(t, \mu) \right)\!(y)\;{ d}\mu(y)\,.
\end{equation*}
   with respect to time, initial data and the model coefficients in bouded Lipschitz distance were proved (see e.g. Theorem 2.11 \cite{Carrillo2012}). 
   These results enabled confirmation both analytically and also in computational experiment of the stability of the particle methods as well as the first order of convergence of these methods (\cite{Carrillo2014}, see Theorem 3.2 and Section 4 for the numerical experiments). For stability results for EBT method see e.g. \cite{Brannstrom2013, jablonskiulikowska}.

Firstly, in the section \ref{sec:model} we introduce the probabilistic model for the seroprevalence data - 
the data describing individuals as having or not been infected in the past. 
In this model we account for the process of data acquisition, including aggregation into cells or subsamples 
characterized by the age and time of test. Both these variables are recorded up to some precision, e.g. one year. The algorithm of sampling from the posterior distribution is 
then introduced, allowing for the aggregation process with an application of pseudo-marginal Monte Carlo Markov Chain (MCMC) \cite{Andrieu2009}. 
The next section is devoted to cohort discretization and relating the posterior distribution obtained with the cohort approach, to the continuous case. 
We show the rate of convergence of the posterior distribution in the cohort case to the 
continuous case with the number of cohorts in the approximation and illustrate this on a simulated dataset. 
Finally, in the last section we apply the method to a real dataset available for varicella in Poland.

\section{Bayesian inference for the model \eqref{eqn0}}\label{sec:model}

\subsection{Data model}
We first describe the seroprevalence data. This type of data characterizes 
individuals who have been tested to establish if they have ever had contact with a disease or not. 
The observations are generally of the form $(Y_{ij},t_{ij},a_{ij})$, where $Y_{ij}$ is a random variable indicating 
whether the person $i$ in sample $j$ has had contact with the disease, at exact test time, $t_{ij}$ and exact age at test, $a_{ij}$. We denote the total number of individuals in the sample $j$ by $N_j$.
Let us assume that:

\[\mathbb P(Y_{ij}=1|t_{ij},a_{ij})=q(t_{ij},a_{ij})\,.\]

The function $q(t,a)$ is the solution of the equation (\ref{eqn0}) supplemented with the boundary condition (\ref{init0}).

The data collection system aggregates data with respect to age and time of testing into subsamples $j$, with some possibility of misclassification. This collection and aggregation process will be represented by the family of functions $\Psi_j$. 
The function $\Psi_{j}(t,a)$ is the probability density function of distribution of time of test and age at test in subsample $j$. We assume that data collection process, at least in short time intervals, was random with respect to test time and age at test, so if no misclassification was present, the $\Psi_{j}$ should be uniform distribution on a product of time and age intervals. However, due to uncertainty of age and time ascertainment it is smoothed on the boundary of the box. 

Let us define $p_j$ as:
\begin{equation}\label{eq:p_with_psi}
p_j= \int_{\mathbb R \times \mathbb R^+ }\Psi_j (t,a) q(t,a){d}t{d}a =\mathbb E_{\Psi}(q) 
\end{equation}
then $Y_j=\sum_{i=1}^{N_j}Y_{ij}$ is distributed according to the binomial distribution $Bin(p_j, N_j)$.
To be able to use standard Bayesian parametric inference we assume that $\lambda(t,a)$ is fully described by a finite dimensional parameter $\theta$. We 
also assume that it is possible to factorise the force of infection: $\lambda(t,a)=\lambda_1(a)\lambda_2(t)$. 
According to the observed data, the incidence (rate of new infections) of many childhood infectious diseases such as varicella is periodic in time. 
Therefore, it seems to be relevant to assume that the function $\lambda_2$  is periodic in time.
In subsequent sections for practical application we will use a function such as $\lambda_2(t)= \sin(\gamma_1t+\gamma_2)+1+\gamma_3$, 
and $\lambda_1(a)$ will be a piecewise constant with values $\alpha_i$, $i=1 ... k$.
 $\theta=(\gamma_1, \gamma_2, \gamma_3, \alpha_1, ..., \alpha_k)\in \mathbb R^+\times \mathbb R \times [0,2\pi) \times (\mathbb R^+)^k$ is than a vector of unknown parameters. 
 In the next part of the paper, we  add indices $\lambda_{\theta}$, $p_{\theta}$ to denote explicitly the dependence on $\theta$.
Next, let us denote the likelihood of the data by $L(\theta|Y)=\prod_j p_{\theta,j}^{Y_j}(1-p_{\theta,j})^{N_j-Y_j}$.
To complete the description of the Bayesian model we need to set a prior distributions on $\theta$, denoted by $f(\theta)$. The posterior distribution is than  proportional to:

\begin{equation}\label{eq:posterior}
\pi(\theta |Y ) \propto 
L(\theta|Y) f(\theta)\,.
\end{equation}

\subsection{Monte Carlo Markov Chain (MCMC) algorithm}

Typically, it is not possible to obtain an analytic form of the joint posterior distribution  and a sample from this distribution 
is obtained by sampling the stationary state of a Markov Chain, for 
which the transition probability distribution depends on the right-hand side of  equation \eqref{eq:posterior}. 
The standard MCMC algorithms, however, require computation of the right-hand side of  equation \eqref{eq:posterior}. Consequently,
in our case, a standard MCMC algorithm cannot be used directly due to the fact that $p_{\theta,j}$ is defined by the integral of the solution to a PDE, 
which typically cannot be computed analytically. 
Therefore to sample from posterior distribution of $\theta$ we use a pseudo-marginal approach \cite{Andrieu2009}. This algorithm still assumes that the solution of the PDE can be computed analytically, but it resolves the integration issue. 
The pseudo-marginal MCMC approach assumes existence of an unbiased, positive estimator of likelihood function, $\hat{L}(\theta|Y)$, which is used to introduce an auxiliary target of form
\begin{equation}\label{eq:aux_target}
\pi(\theta,u)\propto\hat{L}(\theta|Y)f(\theta)p(u)\,,
\end{equation}
where $u$ is a random variable with a distribution $p$ which satisfies 
\[\mathbb{E}[\hat{L}(\theta|Y)]=\int \hat{L}(\theta|Y) p(u) d u=L(\theta|Y)\,.\]
Clearly the marginal distribution of $\theta$ is exactly $\pi(\theta)$. Therefore, if we are able to generate an ergodic Markov chain $\{\theta_n,u_n\}$ 
with stationary distribution $\pi(\theta,u)$ then the sequence $\theta_n$ has the correct stationary distribution. In {\bf Algorithm~\ref{alg:rwm}} we describe the pseudo-marginal 
random walk Metropolis algorithm. Note that the only difference in comparison with the standard random walk Metropolis is that the true likelihood function is replaced by an unbiased estimator.

\begin{algorithm}
\caption{Pseudo-marginal random walk Metropolis}
\label{alg:rwm}
\begin{algorithmic}
\STATE Initialize $\theta_0$ and draw corresponding $\hat{L}(\theta_0|Y)$, where $\hat{L}(\theta|Y)$ is an unbiased, positive estimator of $L(\theta|Y)$ .
\FOR {$n=1$ to $N$}
\STATE Sample proposal $\vartheta\sim\mathcal{N}(\theta_{n-1},\sigma^2\mathbb{Id})$.
\STATE Draw an estimator $\hat{L}(\vartheta|Y)$
\STATE With probability \[
{\rm min}\left\{\frac{\hat{L}(\vartheta|Y)f(\vartheta)}{\hat{L}(\theta_{n-1}|Y)f(\theta_{n-1})}, 1\right\}\;,\]
set $\theta_n=\vartheta$ otherwise $\theta_n=\theta_{n-1}$. 
\ENDFOR
\end{algorithmic}

\end{algorithm}

We propose the following procedure to obtain an unbiased, positive estimator of the likelihood function. 
Consider a sequence of independent random variables $(T_{j,m},A_{j,m})\sim \Psi_j$ for $j=1,\dots,J$ and $m=1,\dots,M$ where $J$ is the 
number of subsamples in the model and $M\geq1$ is an arbitrary  integer.
We define an unbiased estimator of $p_{\theta,j}$ by
\begin{equation}\label{eq:unbiased_p}
\hat{p}_{\theta,j,i}=\frac{1}{M}\sum_{m=1}^M q_\theta(T_{j,m}, A_{j,m})\,,
\end{equation}
for $i=1,\dots,N_j$. Next we define $\hat{L}(\theta|Y)$ by
\begin{equation}\label{eq:defL} \hat{L}(\theta|Y)=\prod_j \prod_{i=1}^{N_j} \hat{p}_{\theta,j,i}^{\mathbf{1}(i\leq Y_j)}(1-\hat{p}_{\theta,j,i})^{\mathbf{1}(i> Y_j)}\,.\end{equation}
Clearly, by construction, $\hat{L}(\theta|Y)$ is positive and $\mathbb{E}[\hat{L}(\theta|Y)]=L(\theta|Y)$. The choice of $M$ is crucial for the efficiency of the pseudo-marginal MCMC. 
Small values of $M$ lead to high variance of $\hat{L}(\theta|Y)$ and consequently poor mixing of the Markov chain. High values of $M$ can lead to  exhaustive computation of
$\hat{L}(\theta|Y)$.
Further, this procedure relies on an explicit solution $q_\theta(t,a)$ for the PDE given by \eqref{eqn0}. 
In the general case when we are not able to compute the solution of the PDE analytically, the solution is obtained by an approximation scheme. 
This leads to an important theoretical question of  stability of the posterior distribution with respect to the approximation. In the next section we show the relevant result for one of the commonly used approximations for the structured population models \cite{Roos2013}. \\
 
\section{Discretization of the PDE constrain equation}\label{sec:cohort} 
Before we present our approach, we  discuss the existing contributions to the Bayesian inverse problems with PDE constraints.
Conjunction of differential equations and data gives rise to a range of inverse problems, 
attracting attention of both applied and theoretical research. In applications the interest often lies in solving the inverse problems given the observed data. 
There are considerable contributions to this field concerning the framing of inverse problems in a Bayesian perspective \cite{Stuart2010, Dashi2016}. The most studied model for  data, $y$, is given by:
\begin{equation}\label{eqn:frame}
y= G(\theta) + \eta\,,
\end{equation}
where $\theta$ is the unknown parameter of interest and $\eta$ is a random variable with mean $0$, 
representing the observational noise. $G$ in turn is the observation's operator, relating the observed quantities to the model. For more complex models, evaluation of $G$ i
nvolves solving  a system of differential equations.
We also assume that additional information is available through the prior distribution $f(\theta)$. 

Well-posedness of infinite dimensional inverse problems depends on both  the properties of the PDE problem and the choice of the prior distribution. Much of the research in the
Bayesian approach to infinite-dimensional inverse problems has been carried out for Gaussian priors. In general, 
the prior has to be chosen so that a function drawn from it is sufficiently regular \cite{Stuart2010, Lasanen2012}. 
Moreover, the practical interest often lies in sampling from the posterior distribution, for which specific algorithms are designed. To be able to implement such algorithms, finite-dimensional approximations are considered. 

The Bayesian approach to inverse problems has been applied to a wide range of problems arising from the models used in physics, geology and atmospheric sciences, see e.g. 
\cite{Lasanen2012a, Cotter2010, Bui-Thanh2013, Dwight2010} and in particular the review paper by Stuart \cite{Stuart2010} and the references where in. These studies are 
usually base on Gaussian priors and also assume Gaussian-noise. Noise does not, however, always follow the Gaussian distribution and 
the wrong noise distribution may lead to poor performance of the estimators. 
Even though specific cases of non-Gaussian noise are considered by Lasanen \cite{Lasanen2012, Lasanen2012a}, 
the analysis is restricted only to the model with additive noise, in the general form of (\ref{eqn:frame}). 
Under certain assumptions, the Helinger distance between the posterior distributions obtained in the approximate problem and the original problem 
is bounded in relation to the error of approximation \cite{Stuart2010, Lasanen2012a, Cotter2010, Bui-Thanh2013, Dwight2010}. For the 
definition and elementary properties of the Hellinger distance we refer to e.g. Definition 6.35 in \cite{Stuart2010}. 

Here we present an application of Bayesian data assimilation to population epidemiological models. Due to the 
limited numbers of individuals in populations, it is unlikely that the observational noise is  follows the Gaussian distribution. 
Rather than using additive Gaussian noise as in (\ref{eqn:frame}), which is appropriate for a model of  measurement of continuous quantities in physics, we model counts as arising from binomial or Poisson distributions, the parameters of which are related to the mathematical models describing the process of population growth and the spread of infection in the population. 
In addition, when dealing with a human population, the data are usually naturally aggregated.
For example, the desired characteristics are measured for an individual of certain age at a specific time point, 
but the age and time are recorded up to some precision, most commonly a year. In effect the available data originate from distributions, the parameters of which are defined by integrals of the solution to the model equation. 
In consequence our data model has a different structure than (\ref{eqn:frame}):
\begin{equation}\label{eqn:frame1}
\sum_{i=1}^n Y_i\sim \text{Bin}(G(\theta),n)\,,
\end{equation} 
where $Y_i$ are binary data and $\text{Bin}(G(\theta),n)$ 
is a Binomial distribution with n trials and probability of success $G(\theta)$. 

Similarly to the results presented above, we define  $G(\theta)$ with the help of an infinite-dimensional partial differential equation problem, the equation (\ref{eqn0}) with 
boundary condition (\ref{init0}). As noted above, this particular problem has  explicit solution. However, if we consider the boundary condition (\ref{init1}), we have to introduce an approximation technique, e.g. using the concept of cohorts as in~EBT.

\subsection{Cohort approach}
When describing the evolution of a population it is often useful to group individuals into cohorts. Such cohorts consist of persons that share a certain feature. 
It is assumed that this feature does not change over time, so that the evolution of a cohort can be followed together. The natural grouping in population studies is by time of 
birth (birth cohorts). If the resolution of the time of birth is high enough then we may expect that the solution for the cohort model approximates the solution for
the continuous model.
First we construct the cohorts by dividing the birth time into a countable set of disjoint intervals, $I_i^\epsilon$, of equal length, $\epsilon$, such that $\mathbb R=\cup_{i=-\infty}^{+\infty}I_i^\epsilon$, $I_i^\epsilon=[x_i^\epsilon-\frac{\epsilon}{2}; x_i^\epsilon+\frac{\epsilon}{2})$ and $x_0^\epsilon=0$, $x_{i+1}^\epsilon=x_i^\epsilon+\epsilon$.

The cohort version of the equation \ref{eqn0} will be a set of ODE's associated with the choice of  $\epsilon$, for all $i \in \mathbb Z$:
\begin{align}\nonumber
a_i^{\epsilon}(t)&=t-x_i^\epsilon \ {\rm for}\ t\geq x_i^\epsilon \\ 
m^{\epsilon}_{\theta,i}(0)&=\int_{I_i^\epsilon}q_{\theta}^\epsilon(t,0)dt\\ \nonumber
\frac{d}{dt} m^{\epsilon}_{\theta,i}(t)&=-\lambda_{\theta}(t,a_i^{\epsilon}(t))m_{\theta,i}^{\epsilon}(t)\ {\rm for}\ t\geq x_i^\epsilon\,.
\end{align}
We define $\hat{q}^{\epsilon}_{\theta}(t,\cdot)=\Sigma_{i\in \mathbb Z} m^{\epsilon}_{\theta,i}(t)\delta_{\{a_i^{\epsilon}(t)\}}(\cdot)$.\\[2ex]
{\it Remark:}
Note that $\hat{q}^{\epsilon}_{\theta}$ is a distributional solution to \eqref{eqn0} with boundary data $\hat{q}_\theta^\epsilon (\cdot,0) = \Sigma_{i\in \mathbb Z} \epsilon\delta_{\{x_i^\epsilon\}}$, which approximates the boundary condition (\ref{init0}), i.e. $q_\theta^\epsilon(t,0)=1$.\\[2ex]
{\it Remark:}
For the boundary condition (\ref{init1}) the system is not explicitly solvable since the boundary condition is dependent on the solution itself. 
In a version of EBT, or in fact very similar in this context splitting method, we use the following approximation (for a full description see \cite{Roos1988,GwiazdaEtAl, Carrillo2012,CaGwUl2013}):
\begin{equation}
m^{\epsilon}_{\theta,i}(0)=\epsilon\sum_{j=1}^{+\infty}\beta(\epsilon \cdot j) m_{\theta,i-j}^\epsilon(x_i^\epsilon)\,.
\end{equation}
In this case we need information on the initial distribution of the age profile of $q$.

\subsection{Stability of posterior probability distribution of $\theta$ with respect to the cohort approximation of the PDE constrain} 
Let us consider the model \eqref{eqn0}. 
We note that the posterior probability distribution of $\theta$ for the exact solution of  equation \eqref{eqn0} , 
$q_{\theta}(t,a) \in L^1(\mathbb{R} \times \mathbb{R}^+)$ and the approximate solution $\hat{q}^{\epsilon}_{\theta}(t,\cdot)=\Sigma_{i\in \mathbb Z} m^{\epsilon}_{\theta,i}(t)\delta_{\{a_i^{\epsilon}(t)\}}(\cdot)$ assuming common prior distribution $f(\theta)$ are described by:

\begin{align} \label{eq:posteriorexact}
\pi(\theta |Y ) &\propto \Pi_j p_{\theta,j}^{Y_j}(1-p_{\theta,j})^{N_j-Y_j} f(\theta)
\\
\label{eq:posteriorcohort}
\hat{\pi}^{\epsilon}(\theta |Y) &\propto \Pi_j (\hat{p}^{\epsilon}_{\theta,j})^{Y_j}(1-\hat{p}^{\epsilon}_{\theta,j})^{N_j-Y_j} f(\theta)\,,
\end{align}
where:

\begin{align*}p_{\theta,j}&=\int_{\mathbb{R}^+\times \mathbb{R}}\Psi_j(t,a)q_{\theta}(t,a)dt da\\
\hat{p}^{\epsilon}_{\theta,j}&= \int_{\mathbb{R}^+}\sum_{i\in \mathbb Z}\Psi_j\Big{(}t,a_i^{\epsilon}(t)\Big{)}m^{\epsilon}_{\theta,i}(t) dt
\end{align*}
and $\theta$ determines the function $\lambda_{\theta}(t,a)$ (the force of infection) and $\Psi_j(t,a)$ describes the data collection and aggregation process. 
The definitions of distances between measures and related  basic facts are presented in \ref{sec:def}. 

\begin{tw}\label{tw1}
In the model \eqref{eqn0} with  boundary condition (\ref{init0}) or (\ref{init1})
let $\Psi_j(t,a) \geq 0$, $\int_{\mathbb{R}^+\times \mathbb{R}^+} \Psi_j(t,a)dtda =1$, $\|\Psi_j\|_{W^{1,\infty}}<C$ for all $j=1 ... J$ and $\theta \in H$, $H\subset \mathbb{R}^n$ a compact set.
Moreover, we assume the following about the set of parameters $H$ and the  function  $\lambda_{\theta}(t,a)$.
\begin{itemize}
\item For every compact set $K \subset [0,+\infty)\times [0,+\infty)$:
\[ \sup_{\theta \in H} \sup_{(a,t)\in K} \lambda_{\theta}(t,a) <+\infty\,.\]
\item For every compact set $K \subset (0,+\infty)\times (0,+\infty)$:
\[\inf _{\theta \in H} \inf_{(t,a)\in K} \lambda_{\theta}(t,a)>0\,.\]
\item $\lambda_{\theta}$ is Lipschitz continuous.
\end{itemize}
then
\[C \cdot W_1\big{(}\pi(\theta|Y),\;\hat{\pi}^{\epsilon}(\theta|Y)\big{)}\leq\|\pi(\theta|Y)-\hat{\pi}^{\epsilon}(\theta|Y)\|_{TV} \leq O(\epsilon)\,,\]
where $W_1$ denotes the Wasserstein distance.
\end{tw}
{\it Remark:}
In the model \eqref{eqn0} with boundary condition (\ref{init0}) we are able to find an explicit solution. However, 
in this simple example we present the method and illustrate the type of structural and regularity conditions needed to guarantee that $p_{\theta,j}$ and $\hat{p}^{\epsilon}_{\theta,j}$ are strictly separated from 0 and 1.\\

\begin{proof}
For simplicity we conduct the full proof only for boundary condition (\ref{init0}). The proof for  boundary condition (\ref{init1}) is analogous. It requires application of the stability result of the approximation in bounded Lipschitz distance presented in \cite{GwiazdaEtAl, Carrillo2012}.\\
Let us consider the measures $\mu$ and $\nu$ defined by:
\begin{align*}
\nu(A)&=\int_A \hat{\pi}^{\epsilon}(\theta)f(\theta) d\theta\\
\mu(A)&=\int_A \pi(\theta)f(\theta) d\theta\,.
\end{align*}
The first inequality is obvious for the compact sets (see \ref{sec:def}). To prove the second inequality, we need to bound the total variation distance between measures $\mu$ and $\nu$. We have
\begin{align*}
\|\mu-\nu\|_{TV}&=\sup_{\|\Phi\|_{\infty}\leq1} \int \Phi(\theta) d(\mu-\nu)(\theta)\\
&\leq\Big{\|}\frac{\hat{\pi}^{\epsilon}(\theta)}{|\nu|}-\frac{\pi(\theta)}{|\mu|}\Big{\|}_{\infty}\\
&\leq \frac{1}{|\mu|}\|\hat{\pi}^{\epsilon}(\theta)-\pi(\theta)\|_{\infty}+\frac{1}{|\mu||\nu|}|\;|\mu|-|\nu| \;| \|\hat{\pi}^{\epsilon}(\theta)\|_{\infty}\\
&\leq\frac{1}{|\mu|} C_{\text{Lip}}|p_{\theta,j}-\hat{p}^{\epsilon}_{\theta,j}|+\frac{1}{|\mu||\nu|}C_{\text{Lip}}|p_{\theta,j}-\hat{p}^{\epsilon}_{\theta,j}|\;,
\end{align*}
where $C_{Lip}$ is the Lipschitz constant of the polynomial defining $\pi(\theta)$ with respect to $p_{\theta,j}$.
On the other hand, 
\[
|p_{\theta,j}-\hat{p}^{\epsilon}_{\theta,j}|=\left|\int_{\mathbb{R}^+\times \mathbb{R}}\Psi_j(t,a)q_{\theta}(t,a)dt  da - \int_{\mathbb{R}^+}\sum_{i\in \mathbb Z}\Psi_j\Big{(}t,a_i^{\epsilon}(t)\Big{)}m^{\epsilon}_{\theta,i}(t)dt\right| 
\,.\]
Note that we can rewrite:
\[\int_{\mathbb{R}}\sum_{i\in \mathbb Z}\Psi_j\Big{(}t,a_i^{\epsilon}(t)\Big{)}m^{\epsilon}_{\theta,i}(t) dt = \int_{\mathbb R^+} \sum_{i\in \mathbb Z}\Psi_j\Big{(}t_i^{\epsilon}(a),a)\Big{)} \tilde{m}^{\epsilon}_{\theta,i}(a) da\,,\]
where $\tilde{m}^{\epsilon}_{\theta,i}(a) =m^{\epsilon}_{\theta,i}(t_i^{\epsilon}(a)) $ and $t_i^{\epsilon}(a) = a-x_i^{\epsilon}$. Additionally for a fixed $a$ we can write:
\[\int_{\mathbb{R}}\Psi_j(t,a)q_{\theta}(t,a)dt = \sum_{i\in \mathbb Z} \int _{l_i^{\epsilon}(a)}^{l_{i+1}^{\epsilon}(a)}\Psi_j(t,a)q_{\theta}(t,a)dt\,,\]
where $l_i^{\epsilon}(a) = a-(x_i^{\epsilon} - \frac{\epsilon}{2}).$ Therefore:
\[|p_{\theta,j}-\hat{p}^{\epsilon}_{\theta,j}| \le \int_{\mathbb R^+}\left|\sum_{i\in \mathbb Z} \int _{l_i^{\epsilon}(a)}^{l_{i+1}^{\epsilon}(a)}\Psi_j(t,a)q_{\theta}(t,a) dt - \sum_{i\in \mathbb Z}\Psi_j\Big{(}t_i^{\epsilon}(a),a)\Big{)} \tilde{m}^{\epsilon}_{\theta,i}(a)\right|da\,.\]
Note that: 
\begin{multline*}
\left|\sum_{i\in \mathbb Z} \int _{l_i^{\epsilon}(a)}^{l_{i+1}^{\epsilon}(a)}\Psi_j(t,a)q_{\theta}(t,a) dt - 
\sum_{i\in \mathbb Z}\Psi_j\Big{(}t_i^{\epsilon}(a),a)\Big{)} \int _{l_i^{\epsilon}(a)}^{l_{i+1}^{\epsilon}(a)}q_{\theta}(t,a)\rm dt\right|\leq\\
{\sum_{\{i \in \mathbb Z: [l_i^{\epsilon}(a),l_{i+1}^{\epsilon}] \cap \text{supp} \Psi_j \}} \text{Lip}_{\Psi_j, \epsilon} 
\int _{l_i^{\epsilon}(a)}^{l_{i+1}^{\epsilon}(a)}q_{\theta}(t,a)dt \le \text{Lip}_{\Psi_j, \epsilon} \;\text{diam}(\text{supp} \Psi_j)\,.}
\end{multline*}
Moreover:
\begin{multline*}
\left|\sum_{i\in \mathbb Z}\Psi_j\Big{(}t_i^{\epsilon}(a),a)\Big{)} \tilde{m}^{\epsilon}_{\theta,i}(a) 
-\sum_{i\in \mathbb Z}\Psi_j\Big{(}t_i^{\epsilon}(a),a)\Big{)} \int _{l_i^{\epsilon}(a)}^{l_{i+1}^{\epsilon}(a)}q_{\theta}(t,a) dt\right| \le\\
\|\Psi_j \|_{\infty} \sum_{\{i \in \mathbb Z: [l_i^{\epsilon}(a),l_{i+1}^{\epsilon}] \cap \text{supp} \Psi_j \}} \left|\tilde{m}^{\epsilon}_{\theta,i}(a) - \int _{l_i^{\epsilon}(a)}^{l_{i+1}^{\epsilon}(a)}q_{\theta}(t,a)dt\right| \le \|\Psi_j \|_{\infty} C \epsilon
\,.\end{multline*}
This last inequality follows from the  Lipschitz continuity of $\lambda_{\theta}$ and the exact formulae for solutions. 

\end{proof}

{\it Remark:}
Note that the approximated solution $\hat{q}^\epsilon_\theta(t,\cdot)$ is defined for cohorts evolving with time. 
Therefore it is not possible to evaluate the solution at an arbitrary point $(T_{j,m},A_{j,m})$ to obtain the estimate provided by  formula \eqref{eq:unbiased_p}. 
It is however possible to sample the time points $T_{j,m}$ from the marginal distribution of $\Psi_j$ and  obtain an unbiased estimate of $\hat{p}^\epsilon_{\theta,j}$ in a similar way to \eqref{eq:unbiased_p} as:
\begin{equation}\label{eq:unbiased_pcohort}
\widehat{\hat{p}^\epsilon}_{\theta,j,i}=\frac{1}{M}\sum_{m=1}^M \sum_{i \in \mathbb Z} \Psi_j(T_{j,m}, a_i^{\epsilon}(T_{j,m}))m^\epsilon_{\theta,i}(T_{j,m})\,.
\end{equation}

\subsection{Numerical tests posterior distribution stability}\label{sec_numtests}
In order to illustrate the rate of convergence of the cohort approximation in the Wasserstein distance,  we selected a simple model with one parameter. 
In order to challenge the algorithm with a model related to the one which we plan to apply to real data in the following section (periodic force of infection) we choose 
the parameter describing the period to be the basis of the toy example. Moreover, the period and the width of the box have been chosen so that the period of the wave is 
twice the width of the box and the sine wave takes the same values for all box borders, this together corresponds to the most challenging case for recognition of the parameter by the cohort approximation.
We constructed the toy example 
as follows.

In equation \eqref{eqn0} supplemented with boundary condition \eqref{init0} we choose the force of infection  $\lambda_\gamma(a,t)=20(\sin(\gamma t)+1.1)$, where $\gamma$ is an 
unknown parameter. The choice of periodic function is motivated by the fact that the incidence of many childhood infections  tend to follow a periodic pattern. 
In this particular case, the solution $q_\gamma(a,t)$ is given by 
\begin{equation}
\label{eq:qtoy}
q_\gamma(a,t)=\exp\left\{-20\left[1.1a-\frac{1}{\gamma}\left[\cos(\gamma t)-\cos\left(\gamma(t-a)\right)\right]\right]\right\}\;.
\end{equation}

We simulate the data assuming $\gamma= \pi \approx 3.14$,
with regularized (piecewise afine) uniform distribution on six rectangles $[0,0.05]\times[0,1],[0,0.05]\times[1,2],\dots,[0,0.05]\times[5,6]$ which corresponds to taking 
{ six subsamples.
Precisely 
\begin{align}
\label{eq:Psi}
\Psi_1(a,t)&=\psi (20a)\psi(t)\\
\Psi_j(a,t)&=\psi(20a)\psi(t-j+1)\nonumber\,,
\end{align}
with
\begin{equation}
\label{eq:psi} 
\psi(x)=\begin{cases}
0 & \text{if $x<-0.01$ or $x>1.01$}\\
50x+0.5 & \text{if $x\geq-0.01$ and $x<0.01$}\\
1& \text{if $x>0.01$ and $x<.99$}\\
50(1-x)+0.5& \text{if $x\geq .99$ and $x<1.01$}\,.
\end{cases}
\end{equation}
Each of the two subsamples consists of $N$ individuals. The subsamples were generated in two steps. 
Firstly, the ages and test times, $(t_{ij},a_{ij})$, $j=1,..., 6$ and $i=1,...,N$ were sampled 
for the individuals in each of the subsamples according to the $\Psi_1\,,...\,,\,\Psi_6$ distributions to reproduce the sampling process. 
Then the result of the test $Y_{ij}$ was sampled with  probability $q(t_{ij},a_{ij})$.
We simulate a dataset with $N=10$ and we set $Unif([0,5])$ as the prior for $\gamma$.

In this toy model we analyse how the error between the posterior distribution 
for the exact solution to PDE, $\pi(\theta|Y)$ defined by equation \eqref{eq:posteriorexact}, 
and the posterior distributions corresponding to the cohort approximations, $\hat{\pi}^{\epsilon}(\theta|Y)$ defined by equation \eqref{eq:posteriorcohort}, 
depends on the number of cohorts. In particular we want to confirm in the numerical experiment that the convergence is of first order, 
which is analytically shown in the Theorem \ref{tw1}.

Next for $N=10$ we run MCMC Algorithm~\ref{alg:rwm} for the true posterior and for cohort approximations with the number of cohorts equal to $2^k$, where $k=0, ..., 6$.
For each case we run a pseudo-marginal RWM of length $10^6$ per each,  with Gaussian proposal with standard deviation $\sigma=.5$ and with $M=500$. 
The chosen proposal corresponds to an acceptance rate of around $10\%$ in all cases.

In figure~\ref{fig:toydens} we present approximations of the density of the unknown parameter 
$\gamma$. For clarity of presentation we omit densities for $2,8,32,64$ cohorts. We observe that the true posterior concentrates around the true value of the parameter. 
The approximation by cohorts for a small number of cohorts is highly biased but converges quickly to the true posterior. For $16$
cohorts, the denisty of  the true posterior and of cohort approximation are almost the same. 
\begin{figure}
\includegraphics[width=\textwidth]{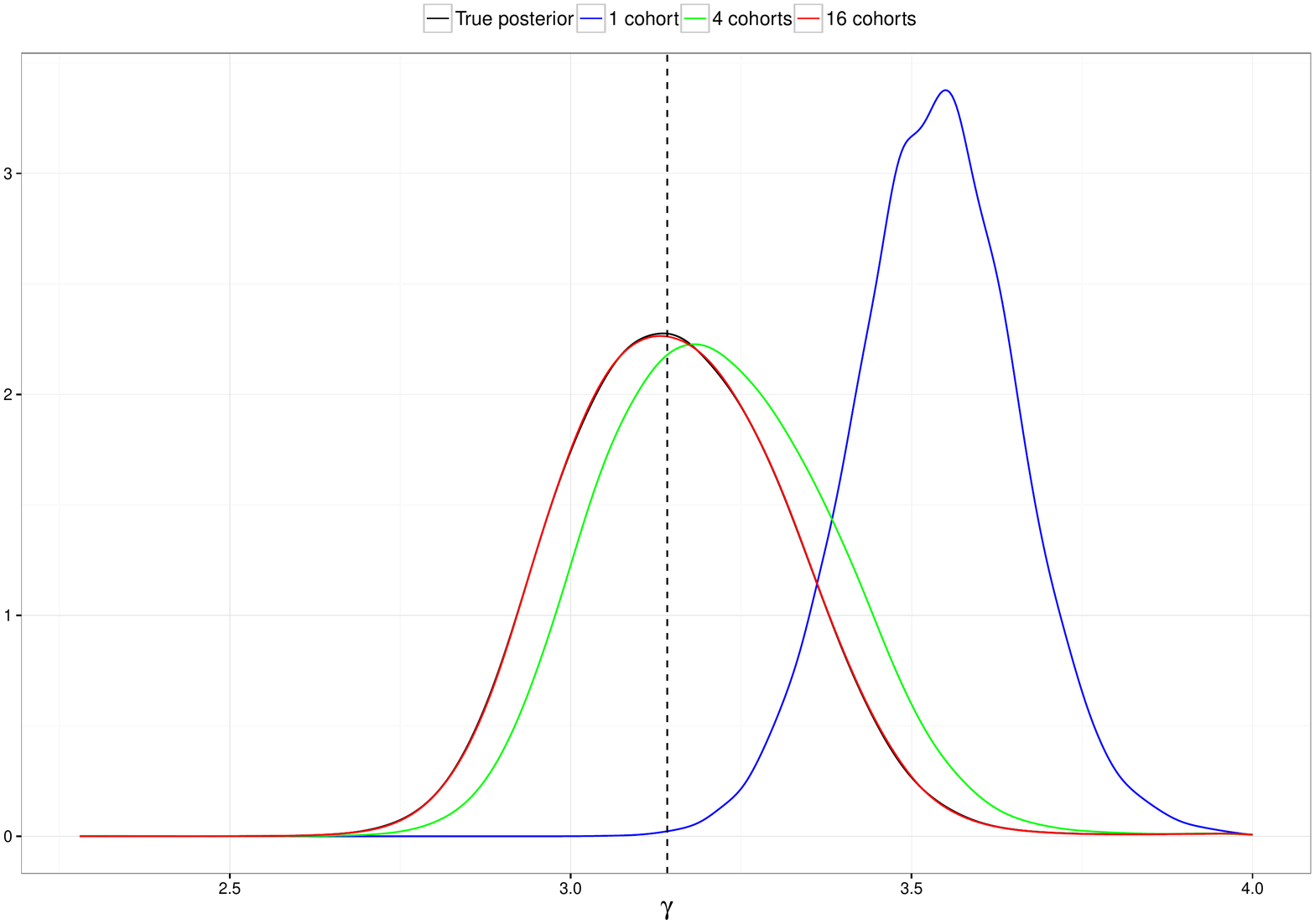}
\caption{Densities of the true posterior and the cohort approximation for the example with $6$ boxes and $N=10$ per box, based on the MCMC algorithm with $10^6$ iterations. The dashed line corresponds to the true value of 
parameter $\gamma=\pi\approx 3.14$.}
\label{fig:toydens}
\end{figure}
We define the error of approximation, $Err(\epsilon)$ as a distance between the reference (exact) solution and the approximated solution at the level of  approximation $\epsilon$. 
The order of convergence is then given by:
\[q:= \lim \limits_{\epsilon \rightarrow 0} \frac{\log (Err(2\epsilon)/Err(\epsilon))}{\log 2}\,.\]
 To estimate the Wasserstein distance between the true posterior and the posterior approximated by cohorts we use
an algorithm from \cite{jablonski2013efficient} applied to empirical measures given by the MCMC algorithm.
Setting  $\epsilon=\frac{1}{2^k}$, for $k\in \mathbb N$, we expect the order of convergence to approximate 1.
The order of the convergence for several elements in this sequence in the Wasserstein metric, for the toy model described above, are given in  Table \ref{tab:wascoh10to6}. 
We note that for $k=6$ the Monte Carlo error (i.e. the error of the numerical approximation of the continuous posterior distribution by the empirical distribution) dominates over the error of approximation by cohorts, which distorts the order of convergence.
 
Moreover, the posterior mean and standard deviation estimators obtained from the cohort approximation are biased, but the bias tends to 0 as the number of cohorts goes to $+\infty$. In 
our toy example this error stabilises for more than 16 cohort, roughly when the Monte Carlo error becomes dominant.

\begin{table}[ht]
\centering
\begin{tabular}{p{2cm}p{2cm}p{2cm}p{2cm}p{2cm}}
\hline
Number of cohorts per box & Wasserstein distance & Order of convergence & Difference of posterior means & Difference of posterior standard deviation \\ 
\hline
1 & 0.394 & -- & 0.369 & 0.167 \\ 
2 & 0.150 & 1.400 & 0.125 & 0.051 \\ 
4 & 0.061 & 1.300 & 0.053 & 0.013 \\ 
8 & 0.022 & 1.470 & 0.019 & 0.007 \\ 
16 & 0.010 &1.170 & 0.001 & 0.003 \\ 
32 & 0.005 &1.020 & 0.002 & 0.010 \\ 
64 & 0.003 &0.902& 0.002 & 0.007 \\ 
\hline
\end{tabular}
\caption{ Comparison of the true posterior for $\gamma=\pi \approx 3.14$ and approximated by cohorts from the data set with $N=10$ observations. All quantities are approximated by median from $5$ independent runs of 
the MCMC algorithm with $10^6$ iteration.}
\label{tab:wascoh10to6}
\end{table}

\section{Application to real data: varicella in Poland}

Varicella or chickenpox is a viral disease which typically occurs in childhood with peak incidence at the age of 4 - 5, when 
children enter preschool or school. In the absence of immunisation programs, the majority of the population contracts the disease by adolescence or early adulthood 
\cite{Bonanni2009}. Once the infection takes place it confers life-time immunity and secondary infections do not generally ocur. 
Vertical infections occur occasionally, when a susceptible mother is infected during pregnancy, but are rare due to the universal immunity among adults. 
The biological marker of past infection exists for this disease, i.e. the presence of antibodies, although this marker may not be ideal due to 
transfer of maternal antibodies to the foetus. The level of maternal antibodies gradually decreases in the child and over 90$\%$ of 
children clear them by the end of the $12$th month of life. Varicella occurs naturally  in short cycles of about 3 - 4 years on top of longer 
cycles of approximately $30$ years as shown for Polish data on Figure \ref{fig:inc}. 
These short fluctuations are generally driven by the accumulation of susceptible individuals and the consequent compensatory epidemic, whereas the long cycle coincides 
with long-term periodic changes of the birth rate resulting in higher or lower proportion of young children in the population. 
The natural cycles are supressed when immunisation programs with substantial coverage are introduced. 
The routine vaccination programs,however, do not currently exist in Poland. The varicella vaccine has beem recommended since 2002, but not performed routinely, resulting in
low uptake and coverage. In particular, before 2008 the coverage was $<5\%$ \cite{biuletyny_szczepien}.

\begin{figure}
\includegraphics[width=\textwidth]{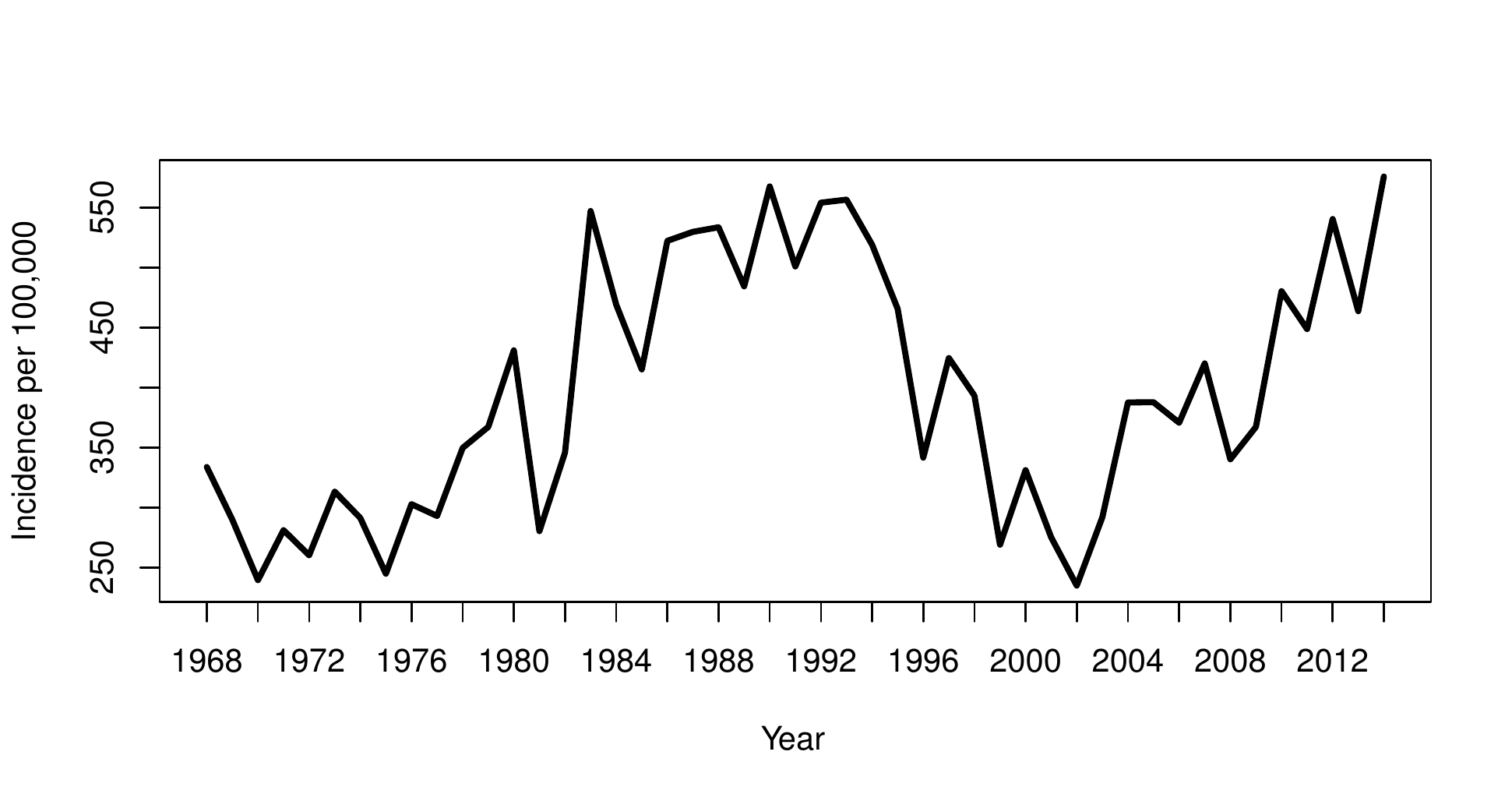}
\caption{ Varicella in Poland. Registered incidence per 100,000 population in 1968 - 2014.}
\label{fig:inc}
\end{figure}

We will demonstrate the use of our method on the sero-prevalence data for varicella in Poland in the time period when the vaccination was uncommon, 
excluding  data from children aged $<12$ months to avoid the potential influence of transferred maternal antibodies. 
The data were derived from the database of the POLYMOD project. Samples from individuals aged 1 - 19 years (by date of birth) at the time of the sample collection 
(2000 - 2004) were extracted from an existing bio-bank and tested for anti-VZV with a commercial testing kit. 
The bio-bank contained samples collected mainly for the purpose of routine check-ups or investigations before  surgical procedures. 
Details of the sample collection and laboratory testing are available elsewhere \cite{Siennicka2009}. 
Altogether 1244 samples were included in the study, the number per year ranged from 108 to 500. The number of individuals in the single $Age \times Year$ cells 
ranged from 1 to 45 and was generally smaller for the 2000 - 2001 period.\\

We consider the proportion of susceptible individual $q(t,a)$ given by \eqref{eqn0} with boundary condition \eqref{init0}. We model the force of infection $\lambda(t,a)$
by
\begin{align}\nonumber
\lambda(t,a)& = \lambda_1(a)(\sin(\gamma_1 t+\gamma_2)+1+\gamma_3)\quad \text{with } \\
\lambda_1(a)&=\sum_{i=1}^k \alpha_i \mathbf{1}(a\in A_i)\,,
\label{eq:force_of_infection}
\end{align}
where $\lambda_1(a)$ is a step function describing the different possible levels of infection in $k$ different age groups $A_i$ of form $A_i=(a_{i-1},a_i]$. 

We choose four groups: children before preschool education $A_1 = (1,3]$, children during preschool education $A_2= (3,7]$, 
primary school students $A_3=(7,15]$, and others $A_4=(15,20]$. 
The force of infection is fully specified by the following unknown parameters: $\alpha_i \in \mathbb{R^+}$ for $ i =1,\dots, 4$, $\gamma_1 \in \mathbb{R^+}$ , $\gamma_2\in[0,2\pi)$ and 
$\gamma_3\in\mathbb{R}^+$. As in section~\ref{sec:model}, we describe seroprevalence data by a binomial Bayesian model. Let $N_{a,t}$ be a number of antibody tests performed during 
the calendar year $t$ for individuals with age $a$ at the time of test, measured as years completed by the time of test, and let
$Y_{a,t}$ corresponding number of positive results. As in section~\ref{sec:model}, we assume that $Y_{a,t}\sim Bin(p_{a,t},N_{a,t})$ with 
\[
p_{a,t}=\int_{[a,a+1)\times[t,t+1)}q(a,t) d a\;  d t\;,
\]
where $q(t,a)$ is the solution of PDE \eqref{eqn0} with boundary condition \eqref{init0} and with the force of infection given by \eqref{eq:force_of_infection}. 
Note that our choice of $\lambda(t,a)$ leads to a closed analytic form of $q(a,t)$.
The solution of PDE \eqref{eqn0} with boundary condition \eqref{init0} with constant level of infection $\alpha$, i.e $\lambda_1(a)\equiv \alpha$, is equal
\[q_{\alpha}(a,t)= \exp\left\{-\alpha\left[ (1+\gamma_3)a+\frac{1}{\gamma1}\left(\cos(\gamma_1t+\gamma_2)-\cos(\gamma_1(t-a)+\gamma_2)\right)\right]\right\}
\,.\]
Hence in our case the function $q(a,t)$ is given by
\[q(a,t)=\mathbf{1}(a\in A_i)q_{\alpha_i}(a,t)\frac{\prod_{j<i}q_{\alpha_{j}}(a_j,t)}{\prod_{j<i}q_{\alpha_{j+1}}(a_j,t)}\;.\]
We choose the distribution $\Psi_j$ as uniform distribution smoothed on boundary, on the unit $Age \times Year$ boxes, see \eqref{eq:Psi}.
We set the following priors: 
\begin{align*}
\alpha_i&\sim Exp(10) \quad \text{for } i=1,\dots,k\\
\gamma_1&\sim Exp(0.8)\\
\gamma_2&\sim Unif([0,2\pi])\\
\gamma_3&\sim Exp(1)\,.
\end{align*}
The choice of hyper-parameters is consistent with prior knowledge on the observed incidence of varicella in Poland as described above. 
Due to the multimodality of the joint posterior distribution, we use a slightly modified adaptive parallel tempering algorithm (APT) 
(introduced in \cite{miasojedow2013adaptive}) for auxiliary target \eqref{eq:aux_target} where only $\hat{L}(\theta|Y)$ is tempered.

The detailed description of the algorithm is given in \ref{sec:appendix}.
We used APT with $5$ levels of temperatures with $5*10^5$ iteration and with burn in time $5*10^4$ and with $M=250$. For each level of temperature we use adaptive scaling RWM with desired acceptance rate equal $10\%$. 
\begin{figure}
\includegraphics[width=\linewidth]{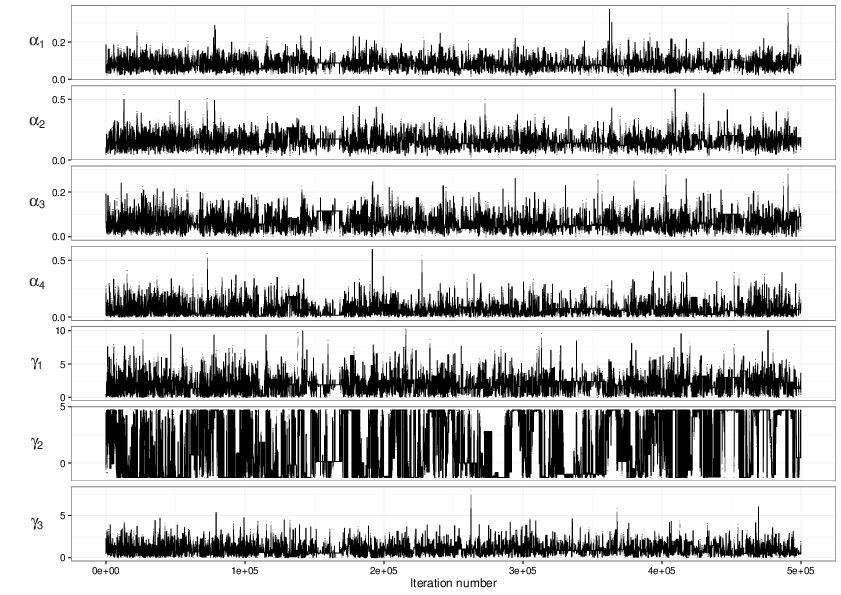}
\includegraphics[width=\linewidth]{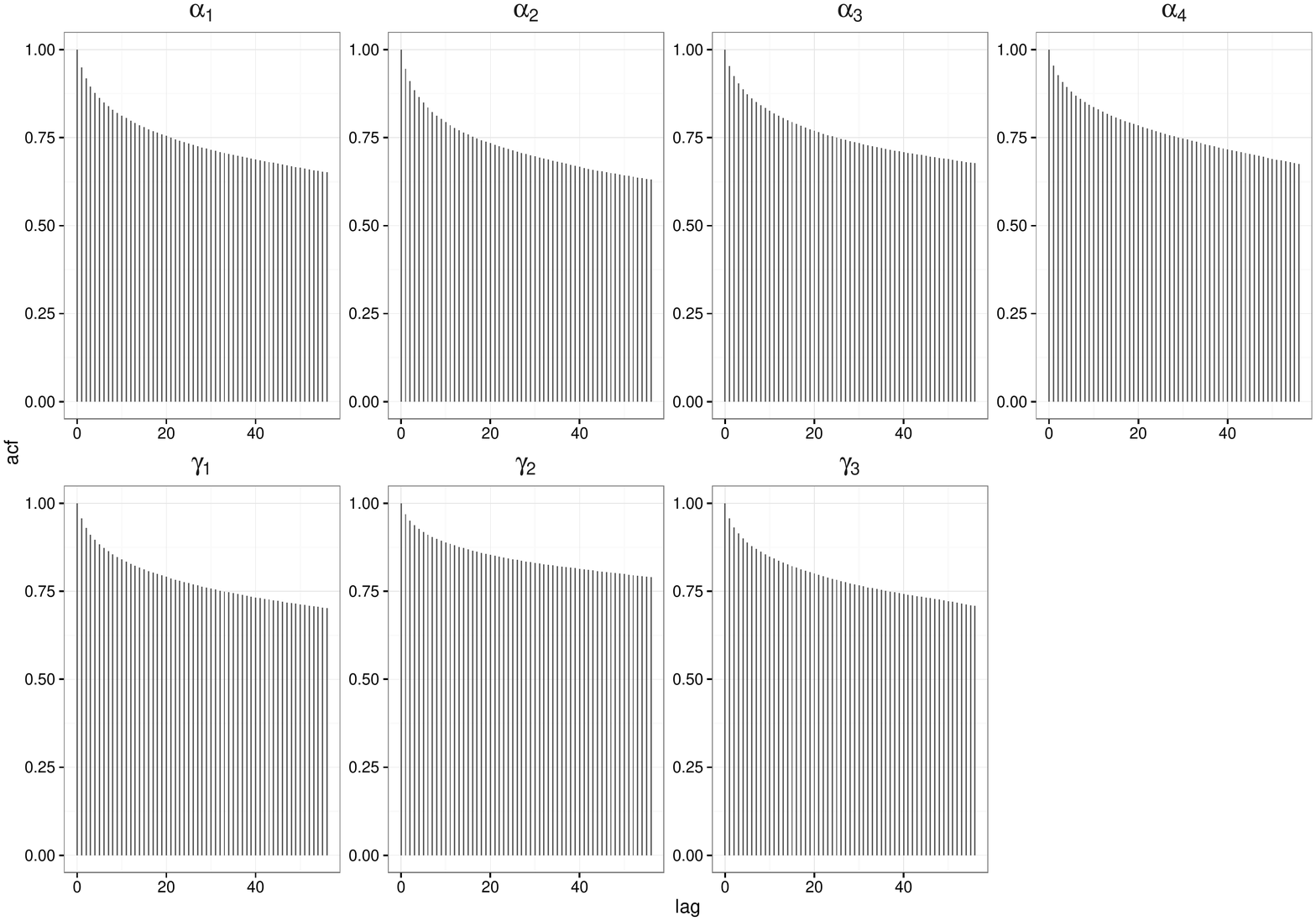}
\caption{Trace plot (top) and autocorelation function (bottom) of A single run of the MCMC algorithm for varicella data set. }
\label{fig:trace}
\end{figure}

We demonstrate the convergence of the algorithm in Figure~\ref{fig:trace}, where we display the trace plots, showing satisfactory 
mixing of the chains and the autocorelation function. This in particular increases the confidence in the shape of the posterior distribution obtained from the model.

\begin{figure}

\includegraphics[width=.9\textwidth]{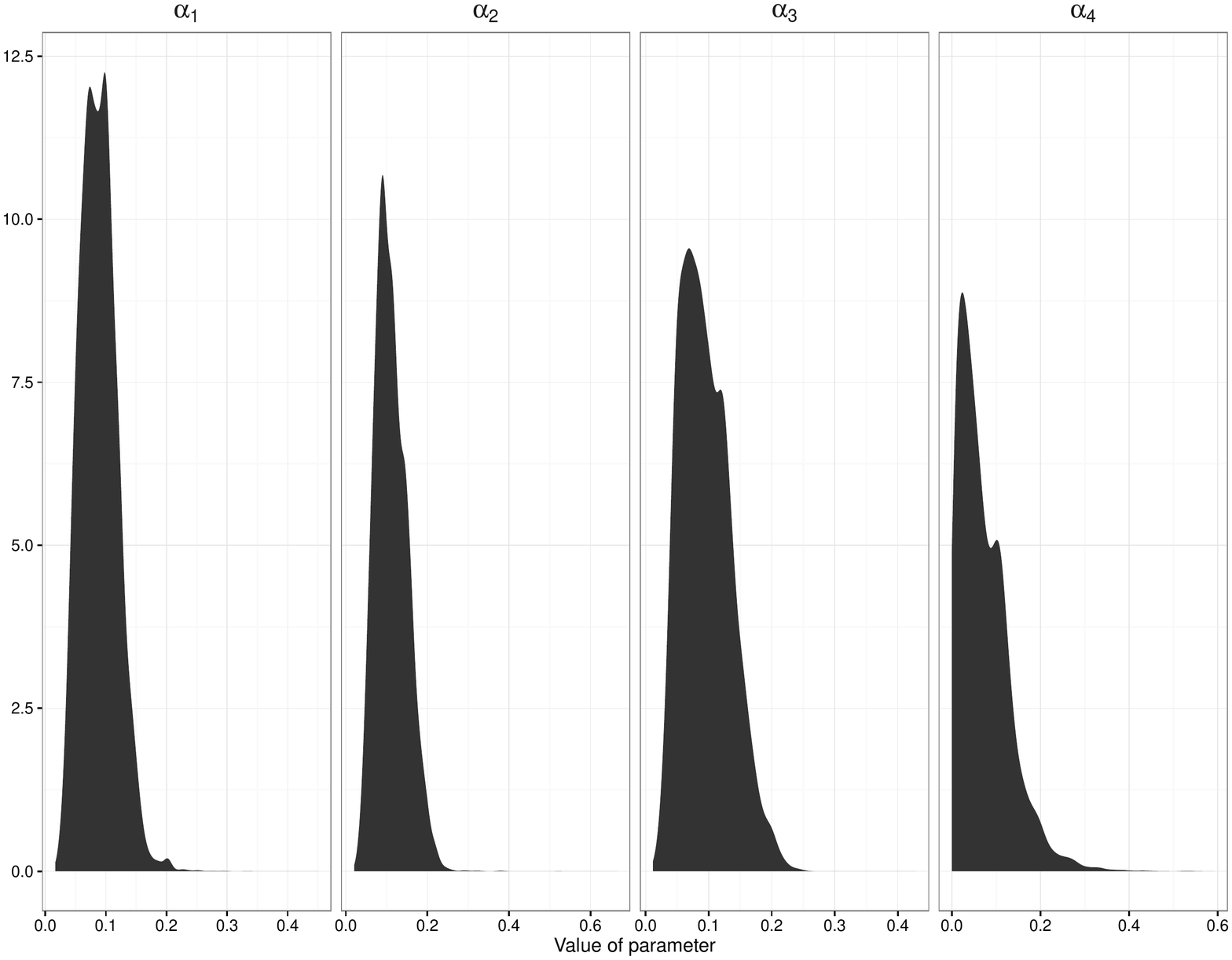}
\includegraphics[width=.9\textwidth]{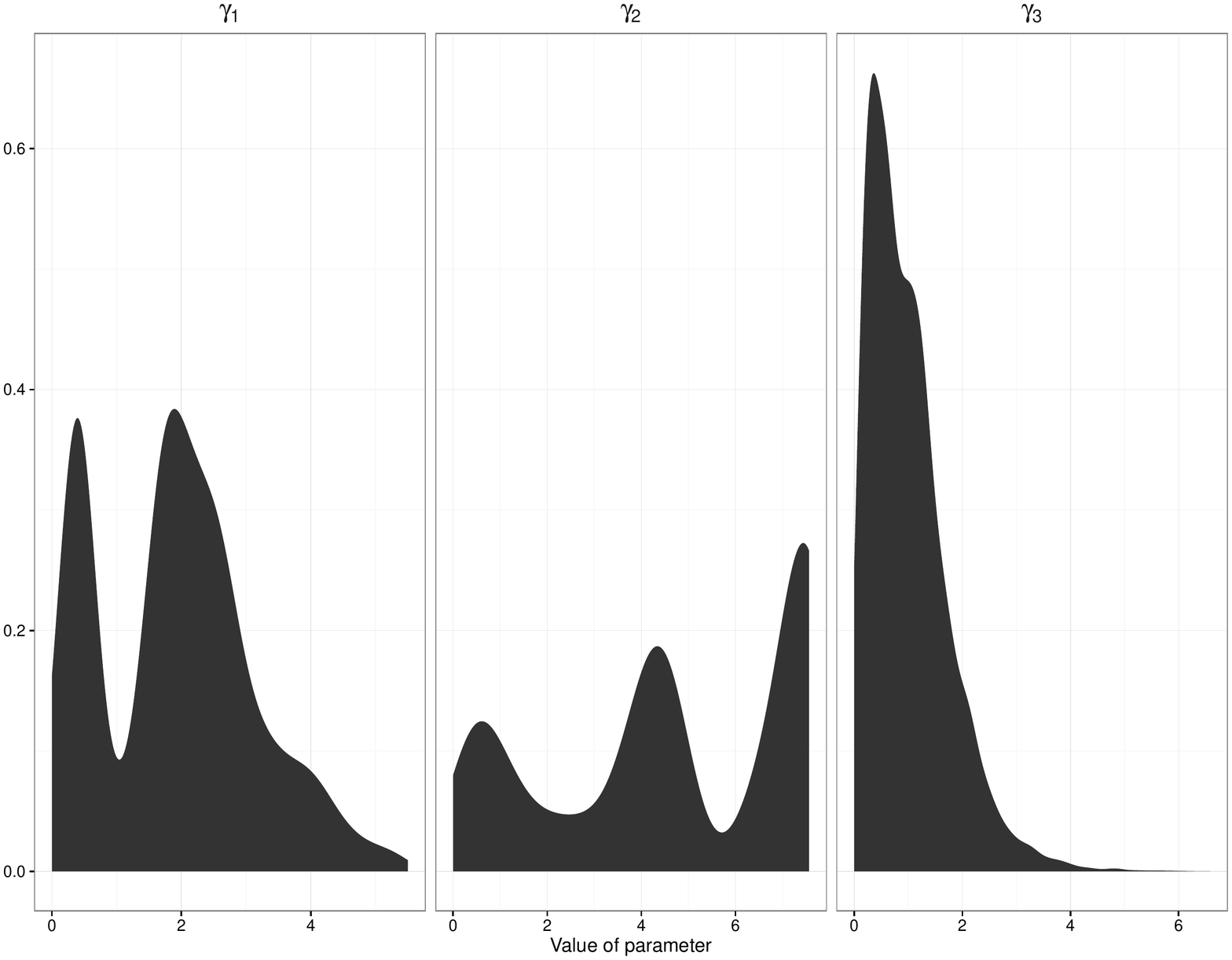}
\caption{Marginal posterior distributions of the parameters. On the top parameters corresponding to the level of the force of infection by age group: $\alpha_1$ - age group $1 - <3$ years; $\alpha_2$ - $3 - <7$ years; $\alpha_3$ - $7-<15$ years; $\alpha_4$- $15 - 19$ years.
On the bottom parameters corresponding to time dependence of the force of infection: $\gamma_1$ - the frequency; $\gamma_2$ - the phase; $\gamma_3>0$ - vertical shift}
\label{fig:dens_alfa}
\end{figure}

The marginal posterior densities of the parameters are given in Figure~\ref{fig:dens_alfa} .
The posterior distributions for the parameters $\alpha_i$ denoting the average level of the force of infection over time in the four age groups demonstrates a 
plausible pattern. The highest value of the posterior mean is observed for the age group 2 ($3 - <7$ years), 0.11, 
markedly higher than for the other groups. This group comprises the pre-school children among whom  peak incidence is usually observed. 
Furthermore, a parameter of interest is the cycle length denoted in our model by $\frac{2\pi}{\gamma_1}$. Interestingly, this distribution is bimodal.
One mode corresponds to the cycle length of $1.9$ years and the other one to a longer cycle, $5.9$ years. This multimodality can appear due to existence of at least two cycles in the observed varicella occurrence, including a long-term cycle.

{ 
Finally, we performed a validation substudy. We repeated the procedure described above, but using only the data from the years $2000-2003$. Next, based on the posterior distributions of the parameters we estimated the age-specific prevalence for the year $2004$. The diagnostic plots for this MCMC are similar to the ones seen for the full dataset and are not shown.
The results of the prediction are presented in  Figure \ref{fig:pred}. The fit of this prediction is very encouraging, with only 2 data points falling outs of the $90\%$ credibility interval.
We also observed that estimators of unknown parameters are almost the same as for the full data set.
}
\begin{figure}
\includegraphics[width=\textwidth]{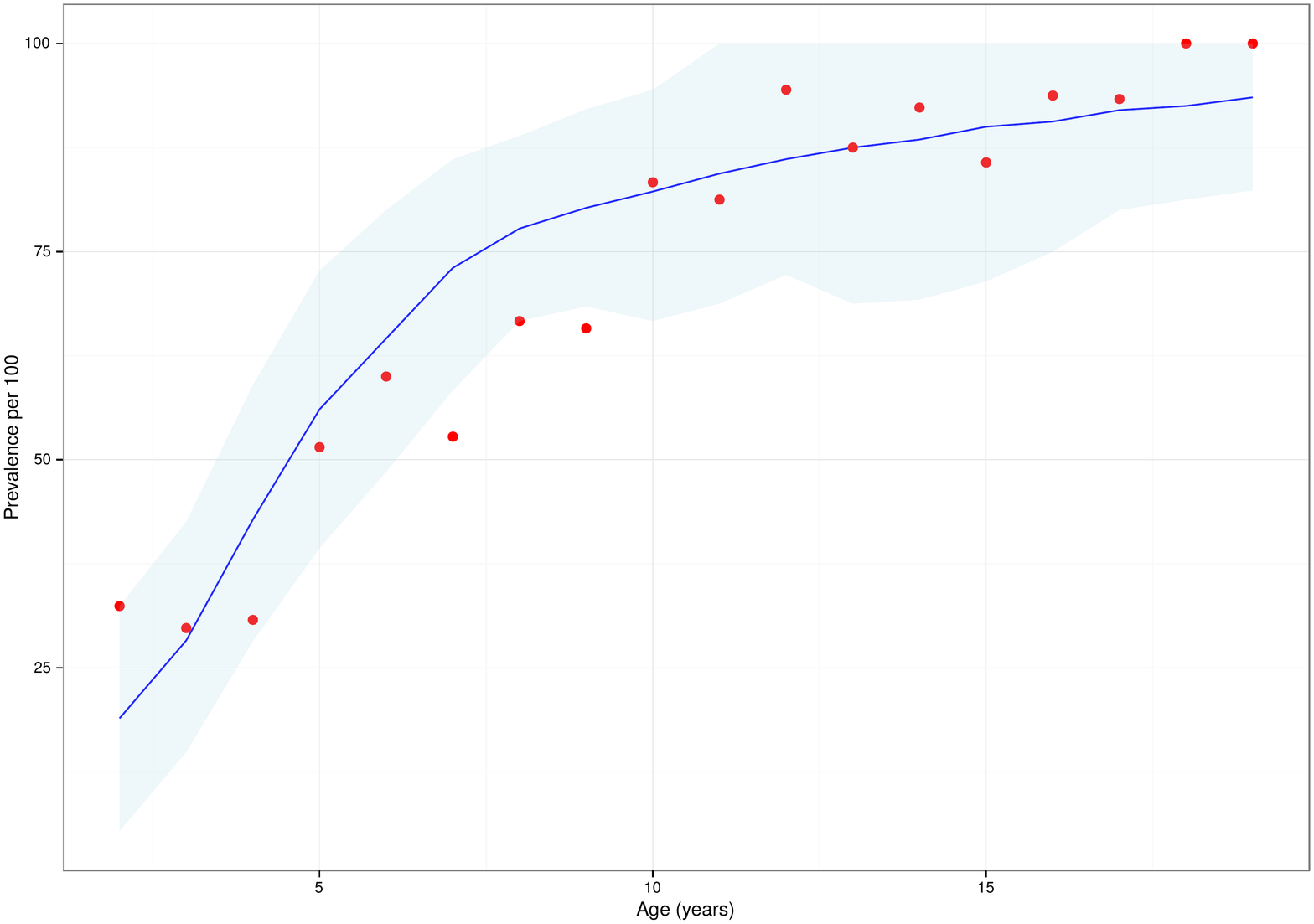}
\caption{Prevalence of anti-VZV antibodies in Poland for year 2004,by age. Red dotes -- the observed values, blue line --the predicted values and the shadowed area -- 90\% credible intervals for the prediction.
Prediction is based on  dataset for years $2000-2003$. }
\label{fig:pred}
\end{figure}

%
%

\section{Concluding remarks}
We note that in previous literature, numerous attempts were made to assimilate seroprevalence data into an equation of the type of  equation \eqref{eqn0}, but the
(lack of) precision of the sampling process was not accounted for (see e.g. a review in \cite{NielHens2012}). 
This corresponds to approximating the integral \eqref{eq:p_with_psi} by a value of $q$ at a single point or on an interval (single cohort). In our toy model developed in  
section \ref{sec_numtests}, we observe that posterior mean of $\gamma$ for a small number of cohorts, and for only a single cohort in particular, may be severely biased, as shown on the Figure \ref{fig:toydens}. 

Our method therefore offers a significant improvement, even for this simple and commonly used model.
What is more, the  methodology presented may be applied to modelling a wide range of different diseases both infectious and non-infectious.
The important feature is the presence of a lasting marker of the disease, which could be measured experimentally. 
We note that the Bayesian approach applied here is 
flexible enough to include additional data sources, for which the data distributions could be parametrised by functions of the parameters defined by the model. 
Examples of such data may include the number of individuals diagnosed by  age of diagnosis and diagnosis time, or data originating from screening programs 
targeting people who have not been diagnosed before. 
We considered here a general form of the force of infection (rate of occurrence of the disease). 
However, for infectious diseases a more complex form describing the disease transmission process could be considered. 
Moreover, additional components can  also be added to the model such as mortality or birth rate as long as the general structure of the models admit the approximation with 
the EBT algorithm or a similar particle approximation. An interesting extension would be to include other relevant structural parameters. 
As an example we may consider a model relating to HIV. 
The HIV population is naturally structured with the status of the immune system, as could be approximated by the CD4 count. 
The existing models usually use the compartmental approach, in which the population members have the same chance of moving into a more advanced stage of disease, 
defined here by the lower ranges of the CD4 count, regardless how long they have in fact spent in the previous stage. 
At the individual level the decline of the CD4 count is a continuous process, so continuous models can potentially produce more accurate estimates. 
In this case the population would be structured in $\mathbb{R}^+ \times \mathbb{R}^+$ to incorporate both age and the CD4 count.

\section*{References}

\bibliography{osparef1}
\appendix
\section{Definitions}\label{sec:def}
We consider a metric space $H\subseteq\mathbb{R}^n$ equipped with the Borel $\sigma$-field $\mathcal{B}(H)$. For any measurable function $\phi\;:\; H\to \mathbb{R}$ for any $1\leq p\leq\infty$
we define the $\Vert\phi\Vert_p$ by
\[
\Vert\phi\Vert_p=
\begin{cases}
\sup_{\theta\in H}|\phi(\theta)& \text{if } p=\infty\\
\left(\int_H |\phi(\theta)|^p\rm d \theta\right)^{1/p}& \text{if } 1\leq p<\infty\;.
\end{cases}
\]
For any Lipschitz function $\phi$ we denote its Lipschitz constant by $\text{Lip}(\phi)$, i.e.
\[ \text{Lip}(\phi)=\sup_{x,y\in H}\frac{|\phi(x)-\phi(y)|}{|x-y|}\;.\]

Let us consider a measures $\mu,\nu$ on $H$, the total variation distance is defined by
\[\Vert \mu -\nu\Vert_{TV}=\sup_{\Vert \phi\Vert_\infty\leq1}\int_H \phi(\theta)\rm d (\mu-\nu)(\theta)\;.\]
Note,  in the case when $\mu$ and $\nu$  admits  densities with respect to Lebesgue measure $f_\mu$ and $f_\nu$, respectivelly the definition of total variation distance is equivalent to
\[\Vert \mu -\nu \Vert_{TV}=\int_H |f_\mu(\theta)-f_\nu(\theta)|\rm d \theta\;.\]
The inequality $\Vert \mu -\nu \Vert_{TV}\leq\int_H |f_\mu(\theta)-f_\nu(\theta)|\rm d \theta$ is obvious, to see the equality  is enough to take the test function $\phi(\theta)=
\mathbf{1}_{\{f_\mu\geq f_\nu\}}(\theta)-\mathbf{1}_{\{f_\mu < f_\nu\}}(\theta)$.

For any $1\leq p <\infty$ we define the Wasserstein distance $W_p(\mu,\nu)$ on a space of probability measures by
\[W_p(\mu,\nu)^p=\inf_{\gamma\in\Gamma(\mu,\nu)}\int_{H\times H} |x-y|^p\rm d \gamma(x,y)\;,\]
where $\Gamma(\mu,\nu)$ is a set of all joint distributions on $H\times H$ with marginals $\mu$ and $\nu$ respectivelly. 
In particular case $p=1$ by the Kantorovich - Rubenstein duality representation $W_1(\mu,\nu)$ is equivalent to
\[
 W_1(\mu,\nu)=\sup_{\text{Lip}(\phi)\leq 1} \int_H \phi \rm d (\mu-\nu)\;.
\]
If $diam(H)<\infty$ then there exists constant $C$ such that for any probabilistic measure $\mu,\nu$ we have
\[W_1(\mu,\nu)\leq C \Vert \mu -\nu\Vert_{TV}\;.\]

We denote by $W^{1,\infty}$ the space of essentially bounded function $\Psi$ with essentially bounded gradient $\nabla \Psi$. By the Rademacher theorem any bounded Lipshitz 
function belongs
to $W^{1\infty}$ and further 
\[\Vert \Psi\Vert_{W^{1,\infty}}=\Vert\Psi\Vert_\infty+\text{Lip}(\Psi)\;.\] 
We also define the bounded Lipshitz distance between measures $\mu,\nu$, also known in the probability theory as the Fortet-Mourier distance, by
\[
 \rho_{BL}(\mu,\nu)=\sup_{\Vert\Psi\Vert_{W^{1,\infty}}\leq 1} \int_H \Psi \rm d (\mu-\nu)\;.
\]
 
\section{Modified adaptive parallel tempering algorithm}\label{sec:appendix}
The parallel tempering algorithm (PT) is MCMC algorithm used in the case when the target distribution is hard to explore, usually due to the multimodality.
The parallel tempering algorithm defines a Markov chain over the product space $\mathcal{X}^L$, where $\mathcal{X}\subseteq\mathbb{R}^d$. Each of the chains $X_k^{(\ell)}$ targets
a `tempered' version $\pi_\ell$ of the
target distribution $\pi$, i.e. $\pi_\ell\propto \pi^{\beta_{\ell}}$. Where $1=\beta_1>\beta_2>\cdots\beta_L$ is a sequnce of inverse temperatures such that the distribution 
$\pi_L$ is easy to explore and adjacent targets $\pi_\ell$ , $\pi_{\ell+1}$ are simmilar. Each time-step may be decomposed into two successive moves: the swap
move and the propagation move. The swap move allows global moves in particular jumps between different modes, the propagation moves locally explore tempered targets at each level.
In our case we wnat to tempere ony likelihood in auxiliary target \eqref{eq:aux_target}. So we define a sequence of targets as follows
\[
 \pi_\ell(\theta,u)\propto \ \hat{L}(\theta|Y)^{\beta_\ell} f(\theta) p(u)\;.
\]
For such choosen target we perform adaptive parallel tempering algorithm proposed in \cite{miasojedow2013adaptive}. The detailed description is given in algorithm \ref{alg:apt}.
For adatation of temperature schedule we use optimal acceptance rate $0.234$ as in \cite{miasojedow2013adaptive} for random walk adaptation we use optimal acceptance rate $0.1$ sugested
by \cite{sherlock2015}.

\begin{algorithm}
\caption{Modified adaptive parallel tempering}
\label{alg:apt}
\begin{algorithmic}
\STATE Initialize $\boldsymbol{\theta}_0=(\theta_0^1,\dots,\theta_0^L)$ and draw corresponding $\hat{L}(\theta_0^1|Y),\dots,\hat{L}(\theta_0^L|Y)$, where $\hat{L}(\theta|Y)$ is an 
unbiased, positive estimator of $L(\theta|Y)$ .
\FOR {$n=1$ to $N$}

\STATE \textbf{Propagation move:}

\FOR{$\ell=1$ to $L$}
\STATE Sample proposal $\vartheta\sim\mathcal{N}(\theta_{n-1}^\ell,\sigma^2_\ell\mathbb{Id})$.
\STATE Draw an estimator $\hat{L}(\vartheta|Y)$
\STATE With probability \[
\alpha^\ell={\rm min}\left\{\frac{\hat{L}(\vartheta|Y)^{\beta_\ell}f(\vartheta)}{\hat{L}(\theta_{n-1}^\ell|Y)^{\beta_\ell}f(\theta_{n-1}^\ell)}, 1\right\}\;,\]
set $\tilde\theta_n^\ell=\vartheta$ otherwise $\tilde\theta_n^\ell=\theta_{n-1}^\ell$. 
\STATE \textbf{Adaptation of standard devation of proposal:}
\STATE  \[\log(\sigma_\ell^{new})=\log(\sigma_\ell^{old})+n^{-0.6}(\alpha^\ell-0.1)\]
\ENDFOR
\STATE \textbf{Swap move:}
\STATE 
\STATE Sample at random $\ell\in\{1,\dots,L-1\}$.
\STATE With probability
\[\eta^\ell=\min\left\{1,\left(\frac{\hat{L}(\tilde\theta^{\ell+1}_n|Y)}{\hat{L}(\tilde\theta^{\ell}_n)|Y}\right)^{\beta_\ell-\beta_{\ell+1}} \right\}\;,
\]
\STATE set $\theta^{\ell}_n=\tilde\theta^{\ell+1}_n$ and $\theta^{\ell+1}_n=\tilde\theta^{\ell}_n$ otherwise $\theta^{\ell}_n=\tilde\theta^{\ell}_n$ and $\theta^{\ell+1}_n=\tilde\theta^{\ell+1}_n$
\STATE For $j\not\in\{\ell,\ell+1\}$ set $\theta^j_n=\tilde\theta_n^j$.  
\STATE \textbf{Adaptation of inverse temperatures:}
\STATE For $\ell=1,\dots,L-1$ denote by  $\rho_\ell=\frac{1}{\beta_{\ell+1}}-\frac{1}{\beta_{\ell}}$  gaps beetween current temperatures.
\STATE  \[\log(\rho_\ell^{new})=\log(\rho_\ell^{old})+n^{-0.6}(\eta^\ell-0.234)\]
\ENDFOR
\end{algorithmic}

\end{algorithm}

\end{document}